\documentclass[12pt]{article}
\usepackage{sigsam, amsmath}

\usepackage{xspace}
\usepackage{footnote}
\usepackage{graphicx}

\newcommand\norm[1]{\lVert#1\rVert}
\newcommand\thet[1]{\theta_{\text{#1}}}

\def\gb{Gr{\"o}bner basis\xspace}

\def\srs{sparse resultant\xspace}

\newcommand{\M}[1]{\mathtt{#1}}
\newcommand{\V}[1]{\mathbf{#1}}

\usepackage{subfig}

\begin{document}

\title{A Novel Application of Polynomial Solvers in\\ mmWave Analog Radio Beamforming}


\author{Snehal~Bhayani$^{\dagger}$,~Praneeth~Susarla$^{\dagger}$,~S.S.~Krishna~Chaitanya~Bulusu$^\ddagger$,\\~Olli~Silven$^\dagger$,~Markku~Juntti$^\ddagger$,~and~Janne~Heikkila$^\dagger$
\\~$\dagger$~Center~for~Machine~Vision~and~Signal~Analysis~(CMVS),\\$\ddagger$~Centre for Wireless Communications (CWC),\\~University~of~Oulu, 90570, Finland.
\vspace{-1cm}
}
%
%
%
%
\date{}
\maketitle
\section{Introduction}
{Beamforming} is a signal processing technique where an array of antenna elements can be steered to transmit and receive radio signals in a specific direction. The usage of millimeter wave ({mmWave}) frequencies and multiple input multiple output ({MIMO}) beamforming are considered as the key innovations of $5^{\mathrm{th}}$ Generation (5G) and beyond communication systems.
The {mmWave} radio waves enable high capacity and directive communication, but suffer from many challenges such as rapid channel variation, blockage effects, atmospheric attenuations, etc.
The technique initially performs beam alignment procedure, followed by data transfer in the aligned directions between the transmitter and the receiver~\cite{5gprotocol_book}. Traditionally, beam alignment involves periodical and exhaustive beam sweeping at both transmitter and the receiver, which is a slow process causing extra communication overhead with {MIMO} and massive {MIMO} radio units. 
In applications such as beam tracking, angular velocity, beam steering etc.~\cite{antennaarrayapprox_2022}, beam alignment procedure is optimized by estimating the beam directions using \textit{first order polynomial approximations.} 
Recent learning-based SOTA strategies~\cite{Learning_beamalignment_2022} for fast mmWave beam alignment also require exploration over exhaustive beam pairs during the training procedure, causing overhead to learning strategies for higher antenna configurations.
Therefore, our goal is to optimize the beam alignment cost functions \textit{e.g.,}~data rate, to reduce the beam sweeping overhead by applying polynomial approximations of its partial derivatives which can then be solved as a system of polynomial equations.
Specifically, we aim to reduce the beam search space by estimating approximate beam directions using the polynomial solvers.
%
%
Here, we assume both transmitter (TX) and receiver (RX) to be equipped with uniform linear array (ULA) configuration, each having only one degree of freedom (d.o.f.) with $N_t$ and $N_r$ antennas, respectively.
\vspace{-0.8cm}
\section{Problem Formulation}\label{sec:problem_formulation}
Let $R=\log_{2}{(1+\frac{\alpha_1 \alpha_2}{\alpha_3}\norm{\mathbf{w}_{\text{rx}}^H \M{H} \mathbf{w}_{\text{tx}}}^2)}$ denote the communication data rate of the mmWave received signal, where $\alpha_1, \alpha_2,\alpha_3 \in \mathbb{R}$ are the known constants, and $\M{H} \in \mathbb{C}^{N_r \times N_t}$ is a matrix of random complex channel values. The matrix $\M{H}$ can be written as $\M{H} = \M{H}_r + j \M{H}_i $, where $j^2 = -1$, and $\M{H}_r, \M{H}_i$ are real $N_r \times N_t$ matrices with known entries. The beamforming vectors $\mathbf{w}_{\text{rx}}$ and $\mathbf{w}_{\text{tx}}$ are functions of the transmitter and receiver beam angles, $\thet{rx}$ and $\thet{tx}$. Altogether, $R$ is considered as a function of $\thet{rx}$, $\thet{tx}$. We formulate the beamalignment problem as to estimate $\thet{rx}$, $\thet{tx}$ by maximizing $R$ given as,
\begin{align}\label{eq:objective_eqn}
    \thet{rx}^*,\thet{tx}^* = \underset{\thet{rx},\thet{tx} \in \mathbb{R}}{\text{argmax}} \ R.
\end{align}
Exploiting the fact that $\thet{rx},\thet{tx} \in \left[ 0, 2 \pi \right]$, one approach is to subdivide the interval into fixed sub-intervals and search for the maxima of $R$ among sub-intervals~\cite{Learning_beamalignment_2022}. However, all the iterative methods require a good starting point. Moreover, it is not possible to know the total number of stationary points for a given function, thus leading to local maxima or saddle points.  Instead, we draw inspiration from computer vision problems, where algebraic methods have gained popularity in recent times. The problems of estimating camera geometry lead to finite systems of polynomial equations which have been successfully solved using the concepts based on the \gb and the \srs~\cite{Kukelova-thesis,larsson2017efficient,MartyushevVP2022,BhayaniKH20}. In this work, we have adopted the \gb-based approach~\cite{larsson2017efficient}.
\vspace{-0.5cm}
\subsection{Algebraic approach for optimization}
The optimization problem in Eq.~\eqref{eq:objective_eqn} can be solved by estimating those points, say $\thet{rx}^*,\thet{tx}^* \in \mathbb{R}$, where the first order partial derivatives of $R$ w.r.t. $\thet{rx}$ and $\thet{tx}$, \textit{i.e.,}~$R_{\thet{rx}} = \frac{\partial R}{\partial \thet{rx}} $ and $ R_{\thet{tx}} = \frac{\partial R}{\partial \thet{tx}}$, vanish. However, $R_{\thet{rx}}$ ($=f_1$) and $R_{\thet{tx}}$ ($=f_2$) are not polynomials, and in order to facilitate an algebraic approach, we {approximate} $f_1$ and $f_2$ as bivariate polynomials. Suppose, $\V{x} = \begin{bmatrix}
      \thet{rx} & \thet{tx}
\end{bmatrix}^\top$. Then, the Taylor series expansions~\cite{rudin86real} of $f_1$ and $f_2$ can be expressed as 
\begin{align}\label{eq:tayl_expn}
     f_1 = \underset{(\V{x}^\V{\alpha},c) \in \mathcal{T}_1}{\sum}  c \V{x}^{\alpha},\
     f_2 = \underset{(\V{x}^\V{\alpha},c) \in \mathcal{T}_2}{\sum} c \V{x}^{\alpha},
 \end{align}
 where $\mathcal{T}_1$ and $\mathcal{T}_2$ denote the sets of coefficient and monomial pairs, occurring in the Taylor series expansion of $f_1$ and $f_2$, respectively.  Suppose $\mathcal{B}_1 $ and $\mathcal{B}_2$ respectively denote the sets of monomials ($\V{x}^\V{\alpha}$) in $\mathcal{T}_1$ and $\mathcal{T}_2$, and $\mathcal{C}_1$ and $\mathcal{C}_2$ respectively denote the sets of coefficients ($c$) in $\mathcal{T}_1$ and $\mathcal{T}_2$. In order to approximate $f_1$ and $f_2$ in Eq.~\eqref{eq:tayl_expn} as polynomials, we truncate the monomial sets $\mathcal{B}_1 $ and $\mathcal{B}_2$ as finite subsets, $\overline{\mathcal{B}_1} \subset \mathcal{B}_1$ and $\overline{\mathcal{B}_2} \subset \mathcal{B}_2$. The corresponding truncated set of coefficients, are $ \overline{C_1} = \lbrace c \in C_1 \mid (\V{x}^\V{\alpha}, c) \in \mathcal{T}_1, \V{x}^\V{\alpha} \in \overline{\mathcal{B}_1} \rbrace $ and $ \overline{C_2} = \lbrace c \in C_2 \mid (\V{x}^\V{\alpha}, c) \in \mathcal{T}_2, \V{x}^\V{\alpha} \in \overline{\mathcal{B}_2} \rbrace.$ Thus, we have approximated $f_1$ and $f_2$ respectively with the polynomials $ p_1 = \underset{c \in \overline{C_1},  \V{\alpha} \in \overline{\mathcal{B}_1}}{ \sum}  c \V{x}^{\alpha}  $ and $
    p_2 =  \underset{c \in \overline{C_2}, \V{\alpha} \in \overline{\mathcal{B}_2}}{ \sum}  c \V{x}^{\alpha}$. The common roots of $p_1$ and $p_2$ represent the approximate solutions to $f_1=f_2=0$. Let the exponent sets of the monomials in $\overline{\mathcal{B}_1}$ and $\overline{\mathcal{B}_2}$ be denoted as $\overline{B_1} \in \mathbb{Z}^2_{\geq 0}$ and $\overline{B_2} \in \mathbb{Z}^2_{\geq 0}$, respectively. Therefore, the functions, $f_1$ and $f_2$, and the truncated polynomials, $p_1$ and $p_2$, can be expressed as
\begin{align}\label{eq:f_and_p}
     f_1 = \underset{(\V{x}^\V{\alpha}, c) \in \mathcal{T}_1}{\underset{\V{\alpha} \in {{B}_1}}{ \sum}}  c \V{x}^{\alpha}, \ & \ 
    f_2 = \underset{(\V{x}^\V{\alpha}, c) \in \mathcal{T}_2}{\underset{\V{\alpha} \in {{B}_2}}{ \sum}}  c \V{x}^{\alpha}, \
     p_1 = \underset{(\V{x}^\V{\alpha}, c) \in \mathcal{T}_1}{\underset{\V{\alpha} \in \overline{{B}_1}}{ \sum}}  c \V{x}^{\alpha}, \ \ 
    p_2 = \underset{(\V{x}^\V{\alpha}, c) \in \mathcal{T}_2}{\underset{\V{\alpha} \in \overline{{B}_2}}{ \sum}}  c \V{x}^{\alpha}.
\end{align}
Here, one of the common roots of $p_1$ and $p_2$ should be as close to a global maxima of $R$ as possible. 

\vspace{3px}
\noindent \textbf{Number of common roots of $p_1$ and $p_2$:} 
The  \textit{Bernstein–Kushnirenko theorem}~\cite{Cox-Little-etal-05} provides the upper bound on the number of common roots of $p_1$ and $p_2$,  denoted as $\eta$, in the complex field $\mathbb{C}^{2}$. Let, $P_i$ denote the convex hull of $\overline{B_i}$, and $\text{Vol}_2 (P_i)$ denote the euclidean volume (area) of $P_i$, for $i=1,2$.  Then, $\eta$ can be considered as a function of the exponent sets, $\overline{B_1}$ and $\overline{B_2}$. Observe that $\eta$ is the upper bound on the size of the matrix undergoing eigenvalue decomposition in the \gb-based polynomial solvers, which in turn affects the speed of the application. We also need to exhaustively iterate through all of the computed roots over the communication channel values (see Sec.~\ref{sec:problem_formulation}), 
and pick the root that corresponds to the largest $R$. Thus, in the interest of application speed, we require $\eta$ be as small as possible. 
\vspace{5px}

\noindent \textbf{Approximation error:} Lowering $\eta$ comes at the cost of accuracy of the solution to the optimization problem in Eq.~\eqref{eq:objective_eqn}, obtained via the roots of polynomial approximations. Let $\thet{rx}^*, \thet{tx}^*$ denote the true maxima of $R$ in Eq.~\eqref{eq:objective_eqn}. Hence, $f_1 $ and $f_2$ vanish at $\thet{rx}^*, \thet{tx}^*$. Also, let $\thet{rx}^\dagger , \thet{tx}^\dagger $ be one of the estimated roots of the polynomials $p_1$ and $p_2$. Then, our objective here is to find $p_1$ and $p_2$, \textit{s.t.,} $\delta = \mid \mid \thet{rx}^\dagger  - \thet{rx}^* \mid \mid_2^2 + \mid \mid \thet{tx}^\dagger  - \thet{tx}^* \mid \mid_2^2 $ is as close to zero as possible\footnote{Note, that $\mid \mid * \mid \mid_2$ denotes the $2$-norm in $\mathbb{R}^2$.}.  One of the ways to investigate the relationship of $\delta$ with $p_1 $ and $p_2$, is by observing the terms we need to drop to obtain $p_1$ and $p_2$ respectively from $f_1$ and $f_2$ in Eq.~\eqref{eq:tayl_expn}. Let $f\mid_{x^*,y^*}$ denote the evaluation of the bivariate function $f$ by assigning the values $x^*$ and $y^*$ to its two variables $x$ and $y$, and let $\delta_{\thet{rx}}=\thet{rx}^\dagger  - \thet{rx}^*$ and $\delta_{\thet{tx}}=\thet{tx}^\dagger  - \thet{tx}^*$. Then, $(f_1-p_1)\mid_{\thet{rx}^\dagger , \thet{tx}^\dagger }$ and $(f_2-p_2)\mid_{\thet{rx}^\dagger , \thet{tx}^\dagger }$ both can be expressed as infinite sums of terms, each term consisting of monomials in $\delta_{\thet{rx}}$ and $\delta_{\thet{tx}}$  as variables. Observe that, $\delta = 0 $ implies that $(f_1-p_1)\mid_{\thet{rx}^\dagger , \thet{tx}^\dagger }$ and $(f_2-p_2)\mid_{\thet{rx}^\dagger , \thet{tx}^\dagger }$ both should vanish, for all possible roots $\thet{rx}^\dagger , \thet{tx}^\dagger$ of $p_1$ and $p_2$. One way to achieve this is by ensuring that the coefficients of the terms in $(f_1-p_1)\mid_{\thet{rx}^\dagger , \thet{tx}^\dagger }$ and $(f_2-p_2)\mid_{\thet{rx}^\dagger , \thet{tx}^\dagger }$ to be as small as possible. In other words, we need to minimize the magnitude of the dropped terms from $f_1$ and $f_2$ in Eq.~\eqref{eq:tayl_expn}. The dropped terms are infinitely many. Instead, we aim to maximize the magnitude of the selected terms in $p_1$ and $p_2$. Thus, we can loosely redefine the approximation error $\delta$ to be the inverse of the sum of the magnitudes of all the coefficients of the terms of  $f_1$ and $f_2$, selected to obtain the approximations $p_1$ and $p_2$. Specifically, $\delta =  \dfrac{1}{\underset{c \in \overline{C_1}}{ \sum}  \lvert c \rvert + \underset{c \in \overline{C_2}}{ \sum}  \lvert c \rvert} $. Our ongoing work seeks to jointly minimize $\eta$ and $\delta$ w.r.t. the  monomial susbets, $\overline{B_1}$ and $\overline{B_2}$. We define this minimization problem as $\mathcal{O} := \underset{\overline{B_1}, \overline{B_2}}{\text{argmin}} \ \eta + \delta$. We then formulate the polynomial approximations $p_1$ and $p_2$ from the solution of the optimization problem $\mathcal{O}$. 
\begin{figure*}[t]
    \centering
    \captionsetup{width=.99\linewidth}
\includegraphics[width=0.99\linewidth]{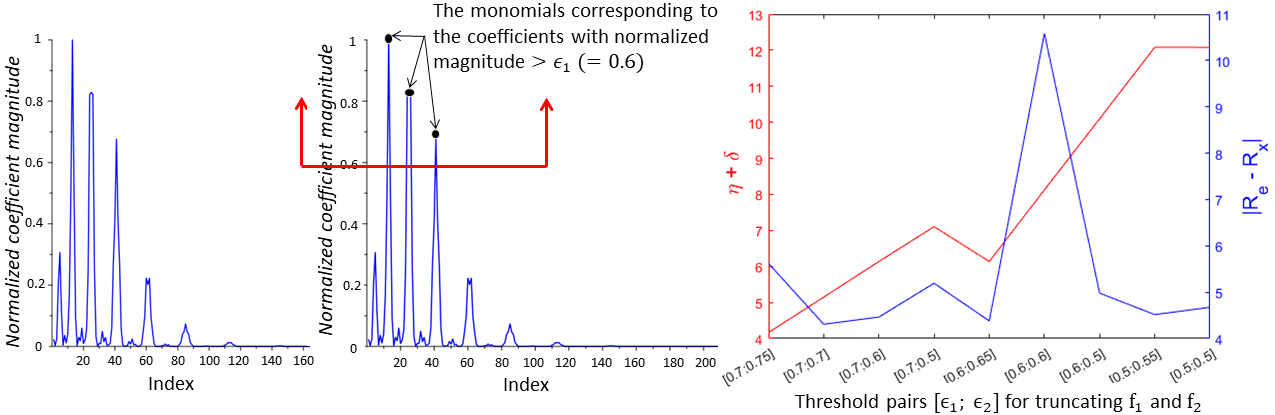}
     \vspace{-5px}

\caption{Let $\V{b}_1$ be a vector of the monomials from $\mathcal{B}_1$ ($\leq$ degree $20$), ordered w.r.t. increasing degree, with the ties broken lexicographically \textit{i.e.,} $\thet{tx} > \thet{rx}$. Let $\V{c}_1$ represent the vector of corresponding coefficient magnitudes (normalized w.r.t. the largest observed magnitude) from $C_1$.  
\textbf{(Left)} A plot displaying $\V{c}_1$; \textbf{(Middle)} The selected indices of the coefficients in $\V{c}_1$ with $\epsilon_1=0.6$; \textbf{(Right)} A plot (red) of the value of the objective function $\eta+\delta$ in the optimization problem $\mathcal{O}$ and a plot (blue) for the absolute difference between the estimated data rate $R_e$ and the known data rate from exhaustive beam search~\cite{Learning_beamalignment_2022} $R_x$, for selected threshold pairs $[\epsilon_1; \epsilon_2]$. 
     } \label{fig:objfn_vs_datarateerr}
     \vspace{-10px}
\end{figure*} 
We note that analytical expressions for $\eta$ and $\delta$ as functions of $\overline{B_1}$ and $\overline{B_2}$ are yet to be determined and are part of our future work. Hence, in this work, 
we adopted a simple strategy to select $\overline{B_1}$ and $\overline{B_2}$, which minimize $\delta$ while keeping $\eta$ \textit{reasonably low}. The strategy is normalize the coefficients in $C_1$ w.r.t.~the largest observed magnitude, and choose $\overline{\mathcal{B}_1}$ corresponding to those coefficients whose normalized values are larger than a certain \textit{threshold}, $\epsilon_1$. We perform the same steps for choosing $\overline{\mathcal{B}_2}$ using $C_2$, via some threshold, $\epsilon_2$. 

\vspace{-3px}
\section{Proposed approach and conclusion}
  In this work, we used a setup of $2 \times 2$ antenna array grid, \textit{i.e.,} $N_t=N_r=2$, and assigned random values to the matrix $\M{H}$,  to demonstrate our approach. We studied some threshold pairs, $[\epsilon_1; \epsilon_2]$ (see the Figure~\ref{fig:objfn_vs_datarateerr} \textbf{(Right)}), and for each pair, we computed the monomial sets $\overline{B_1}$ and $\overline{B_2}$, and solved the corresponding polynomial approximations $p_1$ and $p_2$ using the \gb-based solver~\cite{larsson2017efficient}. For each threshold pair, we also measured the difference in the estimated data rate and the known data rate based on the beam search, and also measured $(\eta + \delta)$ in the optimization problem $\mathcal{O}$. Both these quantities are depicted in Figure~\ref{fig:objfn_vs_datarateerr} w.r.t. the threshold pair. We observe, that the best data-rate estimation (and hence the minimal value of $\delta$) happens when the threshold pair is  $[0.7; 0.7]$. However, the value of the objective function is not minimized ($[0.7; 0.75]$ leads to lower $\eta$ but higher $\delta$), indicating that it came at the expense of $\eta$. Thus, our future work will focus on jointly minimizing, $(\eta + \delta)$.
\vspace{-0.5cm}
\bibliographystyle{IEEEtran}
\bibliography{main}

\end{document}